\documentclass[12pt,a4paper]{article}
\pdfoutput=1
\usepackage{a4wide}
\usepackage[centertags]{amsmath}
\usepackage{amssymb,amsthm}
\usepackage{cite}
\usepackage{tensor}
\usepackage{csquotes}
\usepackage{setspace}
\usepackage[british]{babel}
\usepackage{float}
\usepackage{bbold}
\usepackage[pdftex]{graphicx,color} 
\allowdisplaybreaks
\theoremstyle{definition}
\newtheorem{definition}{Definition}

\begin{document}

\thispagestyle{empty}

\begin{center}
{\Large  \bf On the equivalence of two definitions of conformal primary fields  in $d>2$ dimensions } 
\end{center}

\vspace*{1 cm}

\centerline{Ruben Campos Delgado\footnote{email: ruben.camposdelgado@gmail.com}}
\vspace{1 cm}

\begin{center}{
Bethe Center for Theoretical Physics\\
{\footnotesize and}\\
Physikalisches Institut der Universit\"at Bonn,\\
Nussallee 12, 53115 Bonn, Germany}
\end{center}

\vspace*{1cm}

\centerline{\bf Abstract}
\vskip .3cm
Conformal primary fields are of central importance in a conformal field theory with $d >2$ spacetime dimensions. They can be defined in two ways. A first definition involves commutators between the field and the generators of the conformal group; a second definition characterizes a primary field according to its behavior under a finite conformal transformation. In the existing literature, the proof of the equivalence of the definitions is either omitted or carried out with little details. In this paper we present a clear and concise review of the two definitions and provide a simple and detailed proof for their equivalence, using some minimal results from quantum field theory and basic properties of conformal transformations. The paper is intended as a tutorial for an introductory lecture course in conformal field theory. 
\vskip .3cm

\newpage
\section{Introduction}
Conformal field theory (CFT) is ubiquitous in theoretical physics. It is not a surprise that there exists a vast amount of introductory books and notes aimed to anyone approaching the subject for the first time. A comprehensive introduction is provided by the book of Di Francesco et al. \cite{DiFrancesco:1997nk}. The spacetime dimension is relevant for a CFT. In two dimensions the conformal algebra is infinite-dimensional. In higher dimensions it is of the type $\mathfrak{so}(p,q)$.
Most of the introductory literature focuses on two-dimensional conformal field theories, discussing only marginally the case $d>2$. 
This is motivated, on the one hand, by string theory. There, the CFT arises as a two-dimensional field theory living on the world-volume of a string which moves in some ambient spacetime. Good references specialized in string theory applications are \cite{Polchinski:1998rq, Blumenhagen:2009zz, Ketov:1995yd, Schellekens:1996tg}. A more technical text, well suited to mathematicians, is 
\cite{Schottenloher:2008zz}.
On the other hand, there exist two-dimensional models in statistical mechanics whose continuum description at a second-order phase transition is given by a conformal field theory. A remarkable example is the Ising model for ferromagnets. 
Good references that deal with statistical mechanics applications are \cite{Cardy:2008jc, Ginsparg:1988ui, Dotsenko:1986ca}. 
Further notes which provide a general introduction to conformal field theory, but always with a main focus on two dimensions, are \cite{Qualls:2015qjb, Rovai:2013gga, Efthimiou:2000gz}. 

Nevertheless, conformal field theories in $d>2$ dimensions are also important and do occur in modern research areas like the AdS/CFT correspondence \cite{Maldacena:1997re} or in the context of the conformal bootstrap program \cite{Rattazzi:2008pe}. Some of the introductory references that discuss more deeply conformal field theory in $d>2$ dimensions are indeed the ones originally written as an introduction to the AdS/CFT correspondence \cite{Ammon:2015wua, Nastase:2015wjb, Aharony:1999ti, Zaffaroni:2000vh}. Other sources with pedagogical intents discuss conformal field theories in $d>2$ dimensions employing the projective null cone or embedding space formalism \cite{Osborn:2013ConformalFT, Rychkov:2016iqz, Simmons-Duffin:2016gjk}. 

The fields in a unitary CFT can be characterized by a positive number called conformal dimension. One is generally interested in irreducible representations of the conformal group. In $d>2$ spacetime dimensions there exists indeed a class of fields which not only transform irreducibly under the conformal group but also possess the lowest possible conformal dimension among the other fields of the theory. They are called \textit{conformal primary fields}\footnote{In $d=2$ dimensions there exists a difference between primary and quasi-primary fields. Since we are interested in $d>2$, their analysis falls outside the scope of this paper.}. The concept was originally introduced by Mack and Salam who referred to them as interpolating fields \cite{Mack:1969rr}. In the literature there exist two definitions of conformal primary field. Both of them define it in terms of its behavior under a conformal transformation. The difference is that the first definition is a differential one, it refers to an infinitesimal conformal transformation, while the second refers to a finite one. The first definition involves the commutators between the field, evaluated at $x=0$, and the generators of conformal transformations, in particular dilations and special conformal transformations; the second definition gives an explicit rule to compute the transformed field evaluated at the new point in spacetime. These two kind of definitions are ubiquitous in quantum field theory; one can for example restrict oneself to the rotation group, and classify fields in terms of their behaviour under an infinitesimal or a finite rotation.

Given the central role played by conformal primary fields in a CFT, a coherent discussion of both definitions would be expected. Unfortunately, some textbooks and lecture notes state only one of the two definitions \cite{DiFrancesco:1997nk, Ginsparg:1988ui, Qualls:2015qjb, Rovai:2013gga,Aharony:1999ti, Zaffaroni:2000vh}. Others give them both but they do not show their equivalence \cite{Ammon:2015wua, Nastase:2015wjb}. When this is done, the discussion proceeds rather quickly \cite{Rychkov:2016iqz, Simmons-Duffin:2016gjk}. In another case, one definition is immediately stated while the other emerges only after several chapters dealing with other aspects of the theory \cite{Osborn:2013ConformalFT}.

This paper can be thought of as a useful supplementary material accompanying an introductory lecture course on conformal field theory.  As such, its goal is simple: to give a clear and concise review of both definitions and to provide a simple proof of their equivalence. To that end, we use some minimal results from classical and quantum field theory combined with basic properties of conformal field theory. 
The paper is organised as follows. In Section \ref{sec:review} we review the essential results in field theory that we need for the proof. To be precise, we review the concept of symmetry in classical and quantum field theory, and how it is possible to define a field in terms of its transformation under rotations.  Next, we turn to the definition and basic properties of conformal transformations, and the conformal algebra. The discussion is mainly based on \cite{DiFrancesco:1997nk, Zee:2016fuk} and is not intended to be exhaustive. The reader is referred to the mentioned literature for a complete treatment of the topics. In Section \ref{sec:primaries} we illustrate both definitions of conformal primary fields. In Section \ref{sec:equivalence21} we show how the second definition implies the first one. Section \ref{sec:equivalence12} provides extended calculations of an argument proposed in \cite{Simmons-Duffin:2016gjk}, where it is shown that the first definition implies the second one. 

\section{Basic results in field theory}\label{sec:review}
The content of this section is mainly inspired by Chapters 2 and 4 of the book of Di Francesco et al. \cite{DiFrancesco:1997nk}. Other useful references are \cite{Rychkov:2016iqz, Simmons-Duffin:2016gjk, Poland:2018epd}
\subsection{Symmetries}\label{sec:symmetries}
In classical relativistic field theory, a field $\phi^M$ living in a $d$-dimensional flat spacetime is described by an action 
\begin{equation}
    S=\int d^dx\,\, \mathcal{L}\left(\phi^M, \partial_{\mu}\phi^M \right),
\end{equation}
where $\mathcal{L}$ is the Lagrangian density and $M$ represents some collection of indices. For example, a scalar has $M=\varnothing$;  a rank-2 tensor has $M=\{\mu,\nu\}$. For the sake of brevity we suppress $M$ from now on. In addition, we interchangeably refer to the spacetime position as $x$ or $x^{\mu}$. 
Unless otherwise stated, throughout the whole paper we take the flat metric to be in Minkowski signature, $\eta=\text(-1,+1,...,+1)$.

In general, performing a transformation of coordinates $x\mapsto x'$ has an effect also on the fields:
\begin{equation}\label{eq:trafo_on_fields}
    \phi(x)\mapsto {\phi'}(x')=\mathcal{F}\left(\phi(x)\right),
\end{equation}
where $\mathcal{F}$ is a functional. 
A symmetry is a transformation that leaves the action invariant, $S=S'$.
An important class of symmetries is constituted by continuous global symmetries. They can be written as
\begin{equation}\label{eq:trafo_x_global_cont}
\begin{split}
    x'^{\mu}&=x^{\mu}+\omega^a\frac{\delta x^{\mu}}{\delta\omega_a},\\
    \phi'(x')&=\phi(x)+\omega_a\frac{\delta\mathcal{F}}{\delta\omega_a}(x)
\end{split}
\end{equation}
where $\{\omega_a\}$ is a set of infinitesimal parameters independent of the spacetime position $x$. 
One defines the \textit{generators} $G_a$ of a continuous global symmetry in terms of the difference between the transformed field and the original field, both evaluated at the same position:
\begin{equation}\label{eq:generators_diff}
    \delta_{\omega} \phi(x)= {\phi'}(x)-\phi(x):=-i\omega^a G_a\phi(x).
\end{equation}
In this definition, $G_a$ are differential operators.

The importance of symmetries in physics is summarized by Noether's theorem. It states that to every continuous global symmetry of the action there exists a current which is classically conserved, meaning that 
\begin{equation}
    \partial_{\mu}j^{\mu}_a =0.
\end{equation}
Together with the conserved current there is a conserved charge 
\begin{equation}
    Q_a=\int d^{d-1}x\,\, j^{0}_a
\end{equation}
which is constant in time, $\frac{dQ_a}{dt}=0$. 

A classical field theory can be quantized in several ways, one of them being canonical quantization. In this approach, one promotes fields to operators, $\phi\to \hat{\phi}$, which are required to satisfy commutation or anticommutation relations, depending on their bosonic or fermionic nature. In quantum field theory, one can derive an important relation involving the commutator of the conserved charge and the field (see Section 2.4 of Di Francesco et al. \cite{DiFrancesco:1997nk} or Section 7.3 of Weinberg \cite{Weinberg:1995mt}):
\begin{equation}\label{eq:comm_charge}
    \left[\hat{Q}_a,\hat{\phi}\right]=G_a\hat{\phi}. 
\end{equation}
In other words, the conserved charge $Q_a$, promoted to an operator, is the generator of the symmetry transformation in the Hilbert space of quantum states. 
In order to avoid confusion, we explicitly keep on putting the hat symbol on quantum operators. 

\subsection{Rotations}
Before entering the realm of conformal field theory, it is convenient to step back and recall some known facts about a class of transformations familiar from classical mechanics: rotations. 
Let us suppose we have fixed a Cartesian coordinate system. A rotation $\mathcal{R}$ is a transformation of coordinates that leaves the length of a vector invariant. We can think of it as a matrix, $\vec{v}^{\,\prime}=\mathcal{R}\vec{v}$. We have
\begin{equation}
\vec{v}^{\,\prime 2}=\vec{v}^{\,\prime T}\cdot\vec{v}^{\,\prime}=\left(\mathcal{R}\vec{v}\right)^T\cdot \mathcal{R}\vec{v}=\vec{v}^T \mathcal{R}^T \mathcal{R}\vec{v}\overset{!}{=}\vec{v}^T\cdot\vec{v}.
\end{equation}
The matrices that satisfy the condition $\mathcal{R}^T\mathcal{R}=\mathbb{1}$ are called orthogonal. Recalling that $\text{det}\mathcal{R}_1\mathcal{R}_2=\text{det}\mathcal{R}_1\text{det}\mathcal{R}_2$, their determinant can be either 1 or -1.  The latter case characterizes reflections. Rotations are then a set of matrices with unit determinant that satisfy the orthogonality condition. In $d$ dimensions they make up the special orthogonal group $SO(d)$. It is a Lie group, meaning that a rotation through a finite angle can be obtained by performing infinitesimal rotations repeatedly. An infinitesimal rotation can be written as:
\begin{equation}\label{eq:rot_exp}
    \mathcal{R}=\mathbb{1}+A+O(A^2),
\end{equation}
where $A$ is a matrix proportional to the infinitesimal angle(s). Plugging \eqref{eq:rot_exp} into the orthogonality condition,
\begin{equation}
    \mathcal{R}^T\mathcal{R}=(\mathbb{1}+A)^T(\mathbb{1}+A)+O(A^2)=\mathbb{1},
\end{equation}
gives $A^T=-A$. In $d$ dimensions, there are $d(d-1)/2$ independent matrices $T_a$ that satisfy this condition. These matrices are called generators of rotations, and we can expand $A$ as $A=\sum_a^{d(d-1)/2}\theta_a T_a$, where the real numbers $\theta_a$ are the infinitesimal angles of rotations. 

If we want to consider a finite rotation, we can split the finite angles $\theta_a$ into $N$ pieces so that $\theta_a/N$ is infinitesimal for large $N$, and perform the infinitesimal rotation $N$ times. We then have
\begin{equation}
    \mathcal{R}=\lim_{N\to\infty}\left(1+\sum_{a}\frac{\theta_a T_a}{N}\right)^N=e^{\sum_a\theta_a T_a}
\end{equation}
Of course not only vectors but also fields may be affected by rotations, transforming under some representation of the rotation group. One can even say what a certain kind of field is by looking at its behaviour under rotations. Since $SO(d)$ is a Lie group, it is possible to choose in the definition an infinitesimal or a finite rotation. For example, we can define a scalar field as a function of space that transforms under a finite rotation as $\phi'(x')=\phi(x)$, or $\phi'(x)=\phi\left(\mathcal{R}^{-1}x\right)$. For an infinitesimal rotation, we can expand $\phi'(x)=\phi(x)-A_{ij}x_j\partial_i\phi(x)=\phi(x)-(\theta_a T_a)_{ij}x_j\partial_i\phi(x)$. We can then define a scalar field as a function of space that, under an infinitesimal rotation, undergoes the variation $\delta\phi(x)=-(\theta_a T_a)_{ij}x_j\partial_i\phi(x)$. In particular, for $x=0$, $\delta\phi(0)=0$. If we call $\hat{J}_{ij}$ the quantum charge operator associated to rotational invariance, then in a quantum theory we can state two definitions of a scalar field in terms of its behaviour under rotation:  $\left[\hat{J}_{ij}, \hat{\phi}(0)\right]=0$ or $\hat{\phi'}(x')=\hat{\phi}(x)$.

One can proceed similarly in a conformal field theory. It is possible to define a field by its behaviour under a finite or infinitesimal conformal transformation. This is what we do in Section \ref{sec:primaries}. In a conformal field theory there are however additional subtleties. The conformal group contains not only rotations but also dilations and the so-called special conformal transformations. The goal of the next subsections is to get familiar with them.  

\subsection{Properties of conformal transformations}\label{sec:conformal}
Let $g_{\mu \nu}$ be  the metric tensor in a $d$-dimensional spacetime. A conformal transformation is a change of coordinates $x\mapsto x'=f(x)$ that leaves the metric tensor invariant up to a scale:
\begin{equation}\label{eq:def_conformal}
    g_{\mu \nu}(x) \mapsto g'_{\mu \nu}(x')=\Omega^2(x)g_{\mu \nu}(x),
\end{equation}
where we assume $\Omega(x)>0$. From now on we consider flat spacetime, i.e. $g_{\mu \nu}=\eta_{\mu \nu}$. Recalling the transformation rule for rank-2 tensors with lower indices, we can rewrite Equation \eqref{eq:def_conformal} as 
\begin{equation}\label{eq:defining_conformal_flat}
\frac{\partial x^{\rho}}{\partial {x'}^{\mu}}\frac{\partial x^{\sigma}}{\partial x'^{\nu}}\eta_{\rho \sigma}=\Omega^2(x)\eta_{\mu \nu}.
\end{equation}
Let us now consider an infinitesimal transformation of coordinates
\begin{equation}
    x'^{\rho}=x^{\rho}+\epsilon^{\rho}(x)+O(\epsilon^2).
\end{equation}
Inserting it into Equation \eqref{eq:defining_conformal_flat} gives 
\begin{equation}\label{eq:def_calculations}
\eta_{\mu \nu}-\partial_{\mu}\epsilon_{\nu}-\partial_{\nu}\epsilon_{\mu}+O\left(\epsilon^2\right)=\Omega^2\eta_{\mu \nu}.
\end{equation}
At first order in $\epsilon$,
\begin{equation}\label{eq:k(x)}
    \partial_{\mu}\epsilon_{\nu}+\partial_{\nu}\epsilon_{\mu}=\xi(x)\eta_{\mu \nu},
\end{equation}
for some function $\xi(x)$. Multiplying both sides by $\eta^{\mu \nu}$, using $\eta^{\mu \nu}\eta_{\mu \nu}=d$, fixes  $\xi(x)=\frac{2}{d}\partial \cdot \epsilon$. Hence, Equation \eqref{eq:k(x)} becomes
\begin{equation}\label{eq: equation_conformal_trafo}
    \partial_{\mu}\epsilon_{\nu}+\partial_{\nu}\epsilon_{\mu}=\frac{2}{d}\left(\partial \cdot \epsilon \right)\eta_{\mu \nu}.
\end{equation}
Comparing Equations \eqref{eq: equation_conformal_trafo} and \eqref{eq:def_calculations} we can read off the conformal factor:
\begin{equation}\label{eq:conformal_factor}
    \Omega^2(x)=1-\frac{2}{d}\left(\partial \cdot \epsilon \right) +O\left(\epsilon^2\right).
\end{equation}
The most general solution of equation \eqref{eq: equation_conformal_trafo}, in $d>2$ dimensions, is
\begin{equation}\label{eq:solution}
    \epsilon^{\mu}=a^{\mu} +{\omega^{\mu}}_{\nu}x^{\nu}+\alpha x^{\mu}+2(b\cdot x) x^{\mu} - b^{\mu}x^2,
\end{equation}
where $a^{\mu}$, $\omega^{\mu}_{\nu}$, $b^{\mu}$, $\alpha<<1$  are constants.
The first two terms represent an infinitesimal translation and an infinitesimal Lorentz transformation, respectively. The third term, $x'^{\mu}=(1+\alpha)x^{\mu}$, is a dilation. Its finite version is 
\begin{equation}
    x'^{\mu}=\lambda x^{\mu}.
\end{equation}
The infinitesimal dilation is recovered via $\lambda=1+\alpha +O(\alpha^2)$. 
The last two terms of \eqref{eq:solution} represent a special conformal transformation (SCT). Its finite version is
\begin{equation}\label{eq:sct}
    x'^{\mu}=\frac{x^{\mu}-b^{\mu}x^2}{1-2(b\cdot x)+b^2x^2}.
\end{equation}
A special conformal transformation can be split into more elementary transformations. Equation \eqref{eq:sct} can be indeed written as 
\begin{equation}
    \frac{x'^{\mu}}{x'^2}=\frac{x^{\mu}}{x^2}-b^{\mu}.
\end{equation}
The transformation $x'^{\mu}=\frac{x^\mu}{x^2}=\frac{1}{x_{\mu}}$ is called inversion. A SCT can be then understood as an inversion of $x^{\mu}$, followed by a translation of a factor $b^{\mu}$, and followed again by an inversion. We extensively exploit this property in Section \ref{sec:equivalence21}.

\subsection{The conformal algebra in $d>2$ dimensions}\label{sec:algebra}
The set of conformal transformations forms a group that has the Lorentz group as a subgroup. The algebra of the conformal group is determined by looking at the generators of conformal transformations. The generators, defined as in Equation \eqref{eq:generators_diff}, are
\begin{itemize}
    \item $\mathcal{P}_{\mu}=-i\partial_{\mu}$ for translations;
    \item $\mathcal{J}_{\mu \nu}=i\left(x_{\mu}\partial_{\nu}-x_{\nu}\partial_{\mu}\right)$ for Lorentz transformations;
    \item $\mathcal{D}=-ix^{\mu}\partial_{\mu}$ for dilations;
    \item $\mathcal{K}_{\mu}=-i\left(2x_{\mu}x^{\nu}\partial_{\nu}-x^2\partial_{\mu}\right)$ for special conformal transformations;
\end{itemize}
The non vanishing commutators are 
\begin{equation}\label{eq:conf_algebra}
    \begin{split}
        &\left[\mathcal{D},\mathcal{P}_{\mu}\right]=i\mathcal{P}_{\mu},\\
        &\left[\mathcal{D},\mathcal{K}_{\mu}\right]=-i\mathcal{K}_{\mu},\\
        &\left[K_{\mu},\mathcal{P}_{\nu}\right]=2i\left(\eta_{\mu\nu}\mathcal{D}-\mathcal{J}_{\mu\nu}\right),\\
        &\left[\mathcal{K}_{\rho},\mathcal{J}_{\mu\nu}\right]=i\left(\eta_{\rho\mu}\mathcal{K}_{\nu}-\eta_{\rho\nu}\mathcal{K}_{\mu}\right),\\
       & \left[\mathcal{P}_{\rho},\mathcal{J}_{\mu\nu}\right]=i\left(\eta_{\rho\mu}\mathcal{P}_{\nu}-\eta_{\rho\nu}\mathcal{P}_{\mu}\right),\\
        &\left[\mathcal{J}_{\mu\nu},\mathcal{J}_{\rho\sigma}\right]=i\left(\eta_{\nu\rho}\mathcal{J}_{\mu\sigma}+\eta_{\mu\sigma}\mathcal{J}_{\nu\rho}-\eta_{\mu\rho}\mathcal{J}_{\nu\sigma}-\eta_{\nu\sigma}\mathcal{J}_{\mu\rho}\right),
    \end{split}
\end{equation}
which define the $\mathfrak{so}(d,2)$ algebra, or the $\mathfrak{so}(d+1,1)$ algebra in the case of Euclidean signature. The corresponding conformal group is the Lie group $SO(d+1,1)$ or $SO(d,2)$,  respectively. 

In a quantum theory we also need the charge operators. Therefore, we introduce the operators $\hat{P}_{\mu}$, $\hat{J}_{\mu\nu}$, $\hat{D}$, $\hat{K}_{\mu}$, which are the generators, on the Hilbert space of quantum states, of translations, Lorentz transformations, dilations and special conformal transformations, respectively. They also obey the algebra \eqref{eq:conf_algebra}. The action of the generators on a field can be derived by using the method of induced representations \cite{Mack:1969rr}. One starts by defining the action of the generators on the field at $x=0$:
\begin{equation}\label{eq:commutators_x0}
\begin{split}
    &\left[\hat{J}_{\mu\nu},\hat{\phi}(0)\right]=S_{\mu \nu}\hat{\phi}(0)\\
    &\left[\hat{D},\hat{\phi}(0)\right]=-i\Delta\hat{\phi}(0)\\
    &\left[\hat{K}_{\mu},\hat{\phi}(0)\right]=\kappa_{\mu}\hat{\phi}(0).
\end{split}
\end{equation}
$\Delta$ is a positive fractional number, called conformal dimension; $S_{\mu \nu}$ is a matrix obeying the Lorentz algebra and represents the intrinsic angular momentum (spin) of the field. For instance, scalars have $S_{\mu \nu}=0$; for vectors one can choose ${\left(S^{\mu \nu}\right)^{\rho}}_{\sigma}=i\left(\eta^{\nu \rho}{\delta^{\mu}}_{\sigma}-\eta^{\mu \rho}{\delta^{\nu}}_{\sigma}\right)$; spinor fields have $S_{\mu \nu}=\frac{i}{4}\left[\gamma_{\mu},\gamma_{\nu}\right]$, where the $\gamma_{\mu}$ operators satisfy the $d$-dimensional Clifford algebra
\begin{equation}\label{eq:clifford}
    \{\gamma_{\mu},\gamma_{\nu}\}=2\eta_{\mu \nu}.
\end{equation}
Subsequently, one uses the momentum operator to shift the relations \eqref{eq:commutators_x0} to a generic position $x$. This, together with the commutation relations of the conformal algebra \eqref{eq:conf_algebra}, imply 
\begin{equation}\label{eq:commutators_fields}
\begin{split}
    &\left[\hat{P}_{\mu},\hat{\phi}(x)\right]=\mathcal{P}_{\mu}\hat{\phi}(x)=-i\partial_{\mu}\hat{\phi}(x); \\
    &\left[\hat{J}_{\mu\nu},\hat{\phi}(x)\right]=\mathcal{J}_{\mu\nu}\hat{\phi}(x)=i\left(x_{\mu}\partial_{\nu}-x_{\nu}\partial_{\mu}\right)\hat{\phi}(x)+S_{\mu \nu}\hat{\phi}(x);\\
    &\left[\hat{D},\hat{\phi}(x)\right]=\mathcal{D}\hat{\phi}(x)=\left(-ix^{\nu}\partial_{\nu}-i\Delta\right)\hat{\phi}(x);\\
    &\left[\hat{K}_{\mu},\hat{\phi}(x)\right]=\mathcal{K}_{\mu}\hat{\phi}(x)=\left(-2ix_{\mu}\Delta +ix^2\partial_{\mu}-2ix_{\mu}x^{\rho}\partial_{\rho}-2x^{\nu}S_{\mu \nu}+\kappa_{\mu}\right)\hat{\phi}(x).
\end{split}
\end{equation}

\section{Conformal primary fields in $d>2$ dimensions}\label{sec:primaries}
Loosely speaking, in $d>2$ spacetime dimensions conformal primary fields are irreducible representations of the conformal group with the lowest possible conformal dimension. Since the conformal group has the Lorentz group and the group of dilations as subgroups, every irreducible representation of the conformal group is specified by an irreducible representation $\rho$ of the Lorentz group $SO(d-1,1)$ and a definite conformal dimension $\Delta$. In the special case $d=4$, the Lie algebra of the Lorentz group can be split in terms of two copies of the Lie algebra of $SU(2)$. Hence, in $d=4$ one labels a generic representations of the Lorentz group in terms of two half-integers $(j_L,j_R)$.

In addition, one requires $\kappa_{\mu}=0$, as we want the spectrum of operators to be bounded from below. First, we notice that $\hat{K}_{\mu}$ is a lowering operator with respect to the conformal dimension. Suppose $\lvert\Psi\rangle$ is an eigenstate of $\hat{D}$: $\hat{D}\lvert\Psi\rangle=i\Delta\lvert\Psi\rangle$. Then, using $\left[\hat{D},\hat{K}_{\mu}\right]=-i\hat{K}_{\mu}$,
\begin{equation}
    \hat{D}\hat{K}_{\mu}\lvert\Psi\rangle=\left(\left[\hat{D},\hat{K}_{\mu}\right]+\hat{K}_{\mu}\hat{D}\right)\lvert\Psi\rangle=i(\Delta-1)\hat{K}_{\mu}\lvert\Psi\rangle.
\end{equation}
This allows us to justify the existence of primary operators, considered up to now as an axiom. Suppose we start with any local operator and keep hitting it with $\hat{K}_{\mu}$. Assuming that conformal dimensions are bounded from below, we must eventually hit zero, and this gives us a primary. 

In addition to the boundedness from below of the spectrum, let us suppose now that the theory is also unitary. Then there exists a lower bound for the conformal dimension of the fields. To show this, let us consider the matrix element
\begin{equation}
    A_{\mu{\{t\}},\nu{\{s\}}}={}_{\{t\}}\langle \Delta,l \lvert \hat{K}_{\mu}\hat{P}_{\nu}\rvert\Delta,l\rangle_{\{s\}},
\end{equation}
where $\rvert\Delta,l\rangle$ is a state created by a field operator with dimensions $\Delta$ and spin $l$, and $\{s,t\}$ are spin indices. In a unitary theory, this matrix must have only positive eigenvalues. If it had a negative eigenvalue $\sigma<0$ with eigenvector $\zeta_{\nu,{\{s\}}}$, the state $\lvert\Psi\rangle=\zeta_{\nu,{\{s\}}}\hat{P}_{\nu}\lvert{\{s\}}\rangle$ would have negative norm:
\begin{equation}
    \langle \Psi\lvert\Psi\rangle=\zeta^{\dagger}A\zeta=\sigma\zeta^{\dagger}\zeta<0.
\end{equation}
From the form of the commutator $\left[\hat{K}_{\mu},\hat{P}_{\nu}\right]$ of the algebra \eqref{eq:conf_algebra}, we see that the eigevalues of $A$ gets two contributions. The first is proportional to $\Delta$, the second one corresponds to the eigenvalues of a Hermitian matrix that depends only on the spin:
\begin{equation}
    B_{\mu{\{t\}},\nu{\{s\}}}=\langle{\{t\}} \vert i\hat{J}_{\mu\nu} \rvert\{s\}\rangle.
\end{equation}
The condition that all the eigenvalues of $A$ must be non negative, $\sigma_A\geq0$, translates to $\Delta \geq \sigma_{\text{max}}(B)$, where $\sigma_{\text{max}}(B)$ is the maximum eigenvalue of $B$. A more detailed calculation \cite{Rychkov:2016iqz} shows that, for a symmetric traceless field of spin $l$,
\begin{equation}
    \Delta_{\min}(l)=l+d-2, \hspace{3mm} \text{if} \hspace{3mm} l=1,2,... \hspace{5mm} \text{and} \hspace{3 mm}\Delta_{\text{min}}(0)=\frac{d}{2}-1.
\end{equation}
Analogous results hold for antisymmetric tensors and spinor fields. 

We are now ready to state the first definition.  We reintroduce the set of indices $M$ to give it general validity.
\begin{definition}\label{def:definition1}
{\it In $d>2$ spacetime dimensions, let $\hat{D}$ be the generator of dilations and let $\hat{K}_{\mu}$ be the generator of special conformal transformations. A conformal primary field $\hat{\phi}^M_{\rho}(x)$, in the $\rho$ representation of the Lorentz group and with conformal dimension $\Delta$, satisfies the following conditions at $x=0$: }
\begin{enumerate}
    \item $\left[\hat{D},\hat{\phi}^M_{\rho}(0)\right]=-i\Delta\hat{\phi}^M_{\rho}(0)$;
    \item $\left[\hat{K}_{\mu},\hat{\phi}^M_{\rho}(0)\right]=0$.
\end{enumerate}
\end{definition}
Given a conformal primary, one can construct other operators with higher dimension by acting with the momentum operator $\hat{P}_{\mu}$, which is a raising operator for the conformal dimension (this can be checked as before using $\left[\hat{D},\hat{P}_{\mu}\right]=i\hat{P}_{\mu}$). These new operators are called descendants. 

Conformal primary fields can also be defined via their behaviour under a finite conformal transformation. The ingredient for the second definition is the tensor
\begin{equation}
    {R^{\mu}}_{\lambda}(x)=\Omega^{-1}(x)\frac{\partial x^{\mu}}{\partial x'^{\lambda}},
\end{equation}
where $\Omega(x)$ is the positive square root of the conformal factor of equation \eqref{eq:def_conformal}. It satisfies 
\begin{equation}
    {R^{\mu}}_{\lambda}{R^{\nu}}_{\sigma}\eta_{\mu \nu}=\Omega^{-2}\frac{\partial x^{\mu}}{\partial x'^{\lambda}}\frac{\partial x^{\nu}}{\partial x'^{\sigma}}\eta_{\mu \nu}=\Omega^{-2}\Omega^2\eta_{\lambda\sigma}=\eta_{\lambda \sigma}.
\end{equation}
This means that ${R^{\mu}}_{\rho}$ is a Lorentz transformation. Its explicit expression for an infinitesimal conformal transformation ${x'}^{\mu}=x^{\mu}+\epsilon^{\mu}$, with $\Omega(x)=1-\frac{1}{d}\partial\cdot\epsilon$, is
\begin{equation}\label{eq:Run}
    {R^{\mu}}_{\nu}(x)={\delta^{\mu}}_{\nu}-\frac{1}{2}\left(\partial_{\nu}\epsilon^{\mu}-\partial^{\mu}\epsilon_{\nu}\right).
\end{equation}
We state now the second definition of a conformal primary field.
\begin{definition}\label{def:definition2}
{\it 
In $d>2$ spacetime dimensions, a conformal primary field $\hat{\phi}^M_{\rho}(x)$, in the $\rho$ representation of the Lorentz group and with conformal dimension $\Delta$, transforms under a conformal transformation $\eta_{\mu \nu}\mapsto \Omega^2(x)\eta_{\mu \nu}$ as
\begin{equation}\label{eq:def2}
   \hat{\phi'}^M_{\rho}(x')=\Omega^{\Delta}(x)\mathcal{D}{\left[R(x)\right]^M}_{N}\hat{\phi}^N_{\rho}(x)
\end{equation}
where  ${R^{\mu}}_{\nu}(x)=\Omega^{-1}(x)\frac{\partial x^{\mu}}{\partial x'^{\nu}}$ and $\mathcal{D}{\left[R(x)\right]^M}_{N}$ implements the action of $R$ in the  $SO(d-1,1)$ representation of $\hat{\phi}^{M}_{\rho}(x)$. }
\end{definition}
We now list some possibilities. 
\begin{itemize}
    \item Scalar fields have  $\mathcal{D}\left[R(x)\right]=1$.
    \item Vector fields have $\mathcal{D}{\left[R(x)\right]^{\mu}}_{\nu}={R^{\mu}}_{\nu}(x)$. 
    \item For generic tensor fields a product of $R{^{\mu}}_{\nu}$'s is involved. The position of the indices depends on what kind of tensor $\hat{\phi}^M$ is. For example, if we have ${\hat{\phi}^{\mu \nu}}_{\rho}(x)$, then  ${\mathcal{D}\left[R(x)\right]^{\mu \nu \tau}}_{\lambda \sigma \rho}={R^{\mu}}_{\lambda}(x){R^{\nu}}_{\sigma}(x){R_{\rho}}^{\tau}(x)$.
    \item For spinor fields, $\mathcal{D}\left[R(x)\right]$ is a spinor representation, i.e. $\mathcal{D}\left[R(x)\right]=\exp\left(\frac{i}{2}\Theta_{\mu \nu}S^{\mu \nu}\right)$, where $\Theta_{\mu \nu}$ is an antisymmetric tensor with $d(d-1)/2$ independent components telling us which transformation we are doing, and $S_{\mu \nu} =\frac{i}{2}\left[\gamma_{\mu},\gamma_{\nu}\right]$, with $\gamma_{\mu}$ satisfying the algebra \eqref{eq:clifford}. 
\end{itemize}

The existing introductory textbooks and lecture notes which deal with conformal field theory in $d>2$ present a plethora of possibilities regarding the definitions. 
\begin{itemize}
    \item \cite{DiFrancesco:1997nk, Ginsparg:1988ui, Qualls:2015qjb, Rovai:2013gga} provide only the second definition restricting themselves to scalar fields. Moreover, they adopt the convention of calling the fields  ``quasi-primary'' instead of primary. 
    \item \cite{Aharony:1999ti, Zaffaroni:2000vh} give only the first definition. 
    \item \cite{Ammon:2015wua, Nastase:2015wjb} give both definitions, without explicitly proving the equivalence. 
    \item In \cite{Osborn:2013ConformalFT}  one  definition  is immediately stated while the other emerges only after several chapters dealing with other aspects of the theory.
    \item \cite{Simmons-Duffin:2016gjk} provides both definitions and shows that the first one implies the second one by explicitly using the commutators \eqref{eq:commutators_fields}. Most of the calculations are omitted. 
    \item \cite{Rychkov:2016iqz} provides both definitions and shows that the second one implies the first one. Again, only a quick calculation is provided. 
\end{itemize}

\section{Derivation of the equivalence of the definitions: 2 implies 1}\label{sec:equivalence21} 
In this section we prove that Definition \ref{def:definition2} implies Definition \ref{def:definition1}. We omit the subscript $\rho$ whenever it is clear which type of field we are talking about. We consider a conformal transformation that changes the coordinates $x\mapsto f(x)=x_c$ and the field $\hat{\phi}^M(x)\mapsto{\hat{\phi}^M}_c(x_c)$. We outline here the steps that we are going to follow. 
\begin{enumerate}
    \item We compute the associated conformal factor $\Omega_c(x)$. 
    \item Starting from the expression of ${\hat{\phi}^M}_c(x_c)$ as given in Equation \eqref{eq:def2} we compute ${\hat{\phi}^M}_c(x)$ by noticing that 
    \begin{equation}
    \begin{gathered}
     {\hat{\phi}^M}_c(x_c)=\Omega^{\Delta}(x)\mathcal{D}{\left[R(x)\right]^M}_{N}\hat{\phi}^N(x)=\\
     =\Omega^{\Delta}\left(f^{-1}(x_c)\right)\mathcal{D}{\left[R\left(f^{-1}(x_c)\right)\right]^M}_{N}\hat{\phi}^N\left(f^{-1}(x_c)\right),
     \end{gathered}
\end{equation}
and renaming the indices on both sides we have
\begin{equation}\label{eq:property}
     {\hat{\phi}^M}_c(x)=\Omega^{\Delta}\left(f^{-1}(x)\right)\mathcal{D}{\left[R\left(f^{-1}(x)\right)\right]^M}_{N}\hat{\phi}^N\left(f^{-1}(x)\right).
\end{equation}
    \item We take the difference ${\hat{\phi}^M}_c(x)-{\hat{\phi}^M}(x)$. This, by Equations \eqref{eq:generators_diff} and \eqref{eq:comm_charge}, gives the commutator between the generator of the transformation and the field. At the end we set $x=0$. 
\end{enumerate}
We follow these steps for a special conformal transformation ($c=\text{sct})$ and a dilation ($c=\text{dil}$).
We begin by considering a scalar field, for which $\rho=\mathbb{1}$ (or, in $d=4$, $(j_L,j_R)=(0,0)$) and $\mathcal{D}\left[R(x)\right]=1$. The proof for fields with higher spins is entirely based on the one for scalars. 

\subsection{Step 1: the conformal factors}
Let us first consider a special conformal transformation. Recall that ultimately we want to compute ${\phi^M}_c(x)$. However, if $\Tilde{x}=g(x)$ is the result of a SCT as given in Equation \eqref{eq:sct}, we cannot easily invert $g$ to get $x=g^{-1}(\Tilde{x})$. Therefore, it is more convenient to adopt the interpretation of a SCT as a composition of two inversions and a translation. Let us then compute the conformal factor for an inversion ${x'}^{\mu}=x^{\mu}/x^2$:
\begin{equation}
\begin{gathered}
   \frac{\partial {x'}^{\rho}}{\partial {x}^{\mu}}\frac{\partial {x}'^{\sigma}}{\partial {x}^{\nu}}=\left(\frac{{\delta^{\rho}}_{\mu}}{x^2}-\frac{2x^{\rho}x_{\mu}}{x^4}\right)\left(\frac{{\delta^{\sigma}}_{\nu}}{x^2}-\frac{2x^{\sigma}x_{\nu}}{x^4}\right)= \\
   = \frac{{\delta^{\rho}}_{\mu}{\delta^{\sigma}}_{\nu}}{x^4}-\frac{2{\delta^{\rho}}_{\mu}x^{\sigma}x_{\nu}}{x^4}-\frac{2{\delta^{\sigma}}_{\nu}x^{\rho}x_{\mu}}{x^6}+\frac{4x^{\rho}x^{\sigma}x_{\mu}x_{\nu}}{x^8}.
\end{gathered}
\end{equation}
Contracting with the metric,
\begin{equation}
    \frac{\partial {x'}^{\rho}}{\partial {x}^{\mu}}\frac{\partial {x'}^{\sigma}}{\partial {x}^{\nu}}\eta_{\rho \sigma}=\frac{\eta_{\mu \nu}}{x^4}.
\end{equation}
Comparing this with the defining property of a conformal transformation, equation \eqref{eq:defining_conformal_flat}, we conclude that 
\begin{equation}
   \Omega_{\text{inv}}(x)=x^2.
\end{equation}
Next, let us consider a dilation ${x'}^{\mu}=\lambda x^{\mu}$. We have
\begin{equation}
    \frac{\partial x^{\mu}}{\partial {x'}^{\rho}}\frac{\partial x^{\nu}}{\partial {x'}^{\sigma}}\eta_{\mu \nu}=\frac{{\delta^{\mu}}_{\rho}}{\lambda}\frac{{\delta^{\nu}}_{\sigma}}{\lambda}\eta_{\mu \nu}=\frac{\eta_{\rho \sigma}}{\lambda^2},
\end{equation}
from which 
\begin{equation}
    \Omega_{\text{dil}}(x)=\frac{1}{\lambda}.
\end{equation}
Finally, we notice that for translations the conformal factor is trivially equal to one,  $\Omega_{\text{tr}}(x)=1$. 

\subsection{Step 2: obtaining $\phi_c(x)$ from $\phi_c(x_c)$}
Let $\Tilde{x}$ be the result of an infinitesimal SCT with parameter $b^{\mu}$. We know that $\Tilde{x}^{\mu}/\Tilde{x}^2=x^{\mu}/x^2-b^{\mu}$. In other words,  $\Tilde{x}$ is obtained by performing the following chain of transformations.
\begin{enumerate}
    \item A first inversion $x^{\mu}\mapsto {x'}^{\mu}=f_1\left(x^{\mu}\right)=\frac{x^{\mu}}{x^2}=\frac{1}{x_{\mu}}$. The field transforms as 
    \begin{equation}
        \hat{\phi}(x)\mapsto \hat{\phi}'(x')=x^{2\Delta}\hat{\phi}(x).
    \end{equation}
    \item A translation ${x'}^{\mu}\mapsto {x''}^{\mu}=f_2\left({x'}^{\mu}\right)={x'}^{\mu}-b^{\mu}=\frac{x^{\mu}}{x^2}-b^{\mu}=\frac{\Tilde{x}^{\mu}}{\Tilde{x}^2}$.  The field transforms as
    \begin{equation}
    \hat{\phi}'(x')\mapsto\hat{\phi}''(x'')=1\cdot \hat{\phi}'(x')=x^{2\Delta}\hat{\phi}(x).
    \end{equation}
    \item A second inversion ${x''}^{\mu}\mapsto {x'''}^{\mu}=f_3\left({x''}^{\mu}\right)=\frac{{x''}^{\mu}}{{x''}^2}=\frac{1}{x''_{\mu}}=\Tilde{x}^{\mu}$. The field transforms as
    \begin{equation}\label{eq:final_step}
        \hat{\phi}''(x'')\mapsto\hat{\phi}'''(x''')=\left(x^{''}\right)^{2\Delta}\hat{\phi}^{''}(x'')=\left(x^{''}\right)^{2\Delta}x^{2\Delta}\hat{\phi}(x).
    \end{equation}
\end{enumerate}
We now proceed backwards, starting from $\hat{\phi}'''(x''')$ until we reach $\hat{\phi}'''(x)$. 
Firstly we express $\hat{\phi}'''(x''')$ in terms of $x''$ only. We notice that  $x^{\mu}=\frac{1}{x'_{\mu}}=\frac{1}{x''_{\mu}+b_{\mu}}$. Hence, by Equation \eqref{eq:final_step}, 
\begin{equation}
    \hat{\phi}'''({x'''}^{\mu})=\left({x''}^{\mu}\right)^{2\Delta}\left(\frac{1}{x''_{\mu}+b_{\mu}}\right)^{2\Delta}\hat{\phi}\left(\frac{1}{x''_{\mu}+b_{\mu}}\right).
\end{equation}
We observe that $f^{-1}_3(x''')=\frac{1}{x'''}$. Thus, using \eqref{eq:property} with $f^{-1}_3(x'')=\frac{1}{x''}$ we have
\begin{equation}
    \hat{\phi}'''\left({x''}^{\mu}\right)=\left(\frac{1}{x''_{\mu}}\right)^{2\Delta}\left(\frac{1}{\frac{1}{{x''}^{\mu}}+b_{\mu}}\right)^{2\Delta}\hat{\phi}\left(\frac{1}{\frac{1}{{x''}^{\mu}}+b_{\mu}}\right).
\end{equation}
Secondly, we express $\hat{\phi}'''\left({x''}\right)$ in terms of $x'$ only:
\begin{equation}
    \hat{\phi}'''\left({x''}^{\mu}\right)=\left(\frac{1}{{x'}_{\mu}-b_{\mu}}\right)^{2\Delta}\left(\frac{1}{\frac{1}{{x'}^{\mu}-b^{\mu}}+b_{\mu}}\right)^{2\Delta}\hat{\phi}\left(\frac{1}{\frac{1}{{x'}^{\mu}-b_{\mu}}+b_{\mu}}\right).
\end{equation}
Using \eqref{eq:property} with $f^{-1}_2(x')=x'+b$ we determine
\begin{equation}
    \hat{\phi}'''\left({x'}^{\mu}\right)=\left(\frac{1}{{x'}_{\mu}}\right)^{2\Delta}\left(\frac{1}{\frac{1}{{x'}^{\mu}}+b_{\mu}}\right)^{2\Delta}\hat{\phi}\left(\frac{1}{\frac{1}{{x'}^{\mu}}+b_{\mu}}\right).
\end{equation}
Finally, we express $\hat{\phi}'''\left({x'}\right)$ in terms of $x$ only:
\begin{equation}
    \hat{\phi}'''\left({x'}^{\mu}\right)=\left(\frac{x^2}{{x}_{\mu}}\right)^{2\Delta}\left(\frac{1}{\frac{x^2}{{x}^{\mu}}+b_{\mu}}\right)^{2\Delta}\hat{\phi}\left(\frac{1}{\frac{x^2}{{x'}^{\mu}}+b_{\mu}}\right).
\end{equation}
Using \eqref{eq:property} with $f^{-1}_1(x)=\frac{1}{x}$ we obtain
\begin{equation}\label{eq:final_phi}
    \hat{\phi}'''\left(x^{\mu}\right)\equiv\hat{\phi}_{\text{sct}}(x^{\mu})=\left(\frac{x^{\mu}}{x^2}\right)^{2\Delta}\left(\frac{1}{\frac{x_{\mu}}{x^2}+b_{\mu}}\right)^{2\Delta}\hat{\phi}\left(\frac{1}{\frac{x_{\mu}}{x^2}}+b_{\mu}\right).
\end{equation}
Next, we focus on a dilation ${x}^{\mu}_{\text{dil}}=\lambda x^{\mu}$. Infinitesimally, $\lambda=1+\alpha+O\left(\alpha^2\right)$. According to \eqref{eq:def2} a scalar field transforms as
\begin{equation}
    \hat{\phi}_{\text{dil}}(x_{\text{dil}})=\hat{\phi}_{\text{dil}}(\lambda x)=\left(\frac{1}{1+\alpha}\right)^{\Delta}\hat{\phi}(x),
\end{equation}
from which
\begin{equation}\label{eq:dilation_phi}
    \hat{\phi}_{\text{dil}}(x)=\left(\frac{1}{1+\alpha}\right)^{\Delta}\hat{\phi}\left(\frac{x}{1+\alpha}\right).
\end{equation}

\subsection{Step 3: the action of the generators on the field}
Since $b_{\mu}$ is infinitesimal, we can Taylor expand the factors in Equation \eqref{eq:final_phi} up to first order in $b_{\mu}$. We have
\begin{equation}
\begin{gathered}
    \left(\frac{1}{\frac{x_{\mu}}{x^2}+b_{\mu}}\right)^{2\Delta}=\left[\frac{1}{\frac{x_{\mu}}{x^2}\left(1+b_{\mu}x^{\mu}\right)}\right]^{2\Delta}=\left(\frac{x^2}{x_{\mu}}\right)^{2\Delta}\left(\frac{1}{1+b\cdot x}\right)^{2\Delta}=\\
    =\left(\frac{x^2}{x_{\mu}}\right)^{2\Delta}\left(1-2\Delta b\cdot x\right) + O\left(b^2\right).
\end{gathered}
\end{equation}
Furthermore,
\begin{equation}
    \hat{\phi}\left(\frac{1}{\frac{x_{\mu}}{x^2}+b_{\mu}}\right)=\hat{\phi}\left(\frac{\frac{x^{\mu}}{x^2}+b^{\mu}}{\left(\frac{x_{\mu}}{x^2}+b_{\mu}\right)^2} \right)=\hat{\phi}\left(\frac{\frac{x^{\mu}}{x^2}+b^{\mu}}{\frac{1}{x^2}+\frac{2x\cdot b}{x^2}}\right).
\end{equation}
The argument can be written as
\begin{equation}
\begin{gathered}
    \left(\frac{x^{\mu}}{x^2}+b^{\mu}\right)\left(\frac{1}{x^2}+\frac{2x\cdot b}{x^2}\right)^{-1}=\left(x^{\mu}+x^2b^{\mu}\right)(1-2x\cdot b) + O\left(b^2\right)=\\
    =x^{\mu}-2x^{\mu}(x\cdot b) +x^2 b^{\mu}+O\left(b^2\right).
\end{gathered}
\end{equation}
Thus,
\begin{equation}\label{eq:sct_shifted}
\begin{split}
    \hat{\phi}\left(\frac{1}{\frac{x_{\mu}}{x^2}+b_{\mu}}\right)=\hat{\phi}\left(x^{\mu}-2x^{\mu}(x\cdot b) +x^2 b^{\mu}+O\left(b^2\right)\right)=\\
    =\hat{\phi}\left(x\right)-2(x\cdot b) x^{\mu}\partial_{\mu}\hat{\phi}(x) +x^2b^{\mu}\partial_{\mu}\hat{\phi}(x) +O\left(b^2\right).
\end{split}
\end{equation}
Putting all together,
\begin{equation}
\begin{gathered}\label{eq:final_phippp}
    \hat{\phi}_{\text{sct}}(x)=\left(\frac{x^{\mu}}{x^2}\right)^{2\Delta}\left(\frac{x^2}{x_{\mu}}\right)^{2\Delta}\left(1-2\Delta x\cdot b\right)\big[\hat{\phi}(x)-2(x\cdot b) x^{\mu}\partial_{\mu}\hat{\phi}(x)\\+x^2b^{\mu}\partial_{\mu}\hat{\phi}(x)+
    O\left(b^2\right)\big]= \hat{\phi}(x)-2(x\cdot b) x^{\mu}\partial_{\mu}\hat{\phi}(x)+x^2b^{\mu}\partial_{\mu}\hat{\phi}(x)\\-2\Delta (x\cdot b) \hat{\phi}(x)+O\left(b^2\right).
\end{gathered}
\end{equation}
As explained in Subsection \ref{sec:symmetries}, the differential operators $G_a$ are defined in terms of the difference $\hat{\phi}_{\text{sct}}(x)-\hat{\phi}(x):=-iG_{a}\omega^a\hat{\phi}(x)$. In our case we have, by Equation  \eqref{eq:final_phippp}, 
\begin{equation}\label{eq:first_difference}
    \delta_{b}\hat{\phi}(x)=\hat{\phi}_{\text{sct}}(x)-\hat{\phi}(x)=b^{\mu}\left(-2x_{\mu}x^{\nu}\partial_{\nu}+x^2\partial_{\mu}-2\Delta x_{\mu}\right)\hat{\phi}(x)+O\left(b^2\right).
\end{equation}
If we identify $\omega^a=b^{\mu}$ then, by Equations \eqref{eq:generators_diff} and \eqref{eq:comm_charge}, we have 
\begin{equation}
    \left[\hat{K}_{\mu},\hat{\phi}(x)\right]=\left(-2ix_{\mu}x^{\nu}\partial_{\nu}+ix^2\partial_{\mu}-2i\Delta x_{\mu}\right)\hat{\phi}(x).
\end{equation}
In particular, setting $x=0$ gives
\begin{equation}
     \left[\hat{K}_{\mu},\hat{\phi}(0)\right]=0,
\end{equation}
which is the first requirement of Definition \ref{def:definition1}. 
We can repeat the same procedure for the dilation. Expanding \eqref{eq:dilation_phi} up to first order in $\alpha$,
\begin{equation}
    \hat{\phi}_{\text{dil}}(x)=\left(1-\alpha\Delta\right)\left[\hat{\phi}(x)-\alpha x^{\mu}\partial_{\mu}\hat{\phi}(x)\right]+O\left(\alpha^2\right).
\end{equation}
The difference between the transformed field and the new field is
\begin{equation}
    \delta_{\alpha}\hat{\phi}(x)=\hat{\phi}_{\text{dil}}(x)-\hat{\phi}(x)=\alpha\left(-x^{\mu}\partial_{\mu}-\Delta\right)\hat{\phi}(x) +O\left(\alpha^2\right).
\end{equation}
Hence, 
\begin{equation}
  \left[\hat{D},\hat{\phi}(x)\right] =  -i\left(x^{\mu}\partial_{\mu}+\Delta\right)\hat{\phi}(x). 
\end{equation}  
Setting $x=0$ gives
\begin{equation}
\left[\hat{D},\hat{\phi}(0)\right] =-i\Delta \hat{\phi}(0),
\end{equation}
which is the second requirement of Definition \ref{def:definition1}. 

\subsection{Higher representations}
For fields with higher spin, one needs in principle the explicit expression of ${R^{\mu}}_{\nu}$. In the case of a special conformal transformation, the infinitesimal parameter is 
$\epsilon^{\mu}=2(b\cdot x)x^{\mu}-x^2b^{\mu}$. Thus, by equation \eqref{eq:Run}, 
\begin{equation}
     {R^{\mu}}_{\nu}(x)={\delta^{\mu}}_{\nu}-2\left(x^{\mu}b_{\nu}-b^{\mu}x_{\nu}\right).
\end{equation}
For an infinitesimal dilation with parameter  $\epsilon^{\mu}=\alpha x^{\mu}$ the ${R^{\mu}}_{\nu}$ tensor is trivial:
\begin{equation}
    {R^{\mu}}_{\nu}(x)={\delta^{\mu}}_{\nu}-\frac{1}{2}{\delta^{\mu}}_{\nu}\alpha+\frac{1}{2}{\delta^{\mu}}_{\nu}\alpha={\delta^{\mu}}_{\nu}.
\end{equation} 
We can summarize both possibilities by saying that $R$ is equal to the identity plus, eventually, terms directly proportional to $x$. Any representation $\mathcal{D}\left[R(x)\right]$ must replicate this structure. Since at the end we are interested in $x=0$, $\mathcal{D}\left[R(x)\right]$ can be considered trivial and no additional features are introduced to the calculations previously carried out for scalars. However, let us be painstaking and do the calculations anyway to convince ourselves. We consider a vector field 
$\phi^{\mu}$. According to \eqref{eq:def2} it transforms, under a finite conformal transformation, as 
\begin{equation}
 {\hat{\phi}}^{\prime\mu}\left(x'\right)=\Omega^{\Delta}(x){R^{\mu}}_{\nu}(x){\hat{\phi}}^{\nu}(x). 
\end{equation}
The overall procedure is similar to the scalar case. The only difference is that we have to evaluate ${R^{\mu}}_{\nu}(x)$ in the shifted position $x^{\mu}-2x^{\mu}(x\cdot b)+x^2b^{\mu}$, as we did for $\hat{\phi}(x)$ in Equation \eqref{eq:sct_shifted}. Up to first order, the original function and the shifted one coincide:
\begin{equation}
{R^{\mu}}_{\nu}\left(x-2x(x\cdot b)+x^2b\right)={\delta^{\mu}}_{\nu}-2\left(x^{\mu}b_{\nu}-b^{\mu}x_{\nu}\right)+O\left(b^2\right)={R^{\mu}}_{\nu}(x)+O\left(b^2\right).
\end{equation}
Hence, Equation \eqref{eq:final_phippp} generalizes to
\begin{equation}
\begin{gathered}
    {\hat{\phi}_{\text{sct}}}^{\mu}(x)=\\(1-2\Delta x\cdot b)\left[{\delta^{\mu}}_{\nu}-2\left(x^{\mu}b_{\nu}-b^{\mu}x_{\nu}\right)\right]\left[1-2(x\cdot b)x^{\mu}\partial_{\mu}+x^2b^{\mu}\partial_{\mu}\right]\hat{\phi}^{\nu}(x)
    \\+O\left(b^2\right)=\hat{\phi}^{\mu}(x)-2(x\cdot b)x^{\rho}\partial_{\rho}\hat{\phi}^{\mu}+x^2b^{\rho}\partial_{\rho}\hat{\phi}^{\mu}-2\Delta(x\cdot b)\hat{\phi}^{\mu}+\\
  +2b^{\mu}x_{\nu}\hat{\phi}^{\nu} -2x^{\mu}b_{\nu}\hat{\phi}^{\nu}+O\left(b^2\right).
\end{gathered}
\end{equation}
When $x=0$ this implies that
\begin{equation}
    \left[\hat{K}^{\nu}, \hat{\phi}^{\mu}(0)\right]=0.
\end{equation}
Since ${R^{\mu}}_{\nu}$ is already trivial for dilations, the calculations are  identical to the scalar case without any additional modification and lead to 
\begin{equation}
\left[\hat{D},\hat{\phi}^{\mu}(0)\right]=-i\Delta\hat{\phi}^{\mu}(0).
\end{equation}
Similar arguments apply for a generic tensor field of type $(p,q)$. According to \eqref{eq:def2} it transforms as 
\begin{equation}
  {{\hat{\phi}}^{\prime\mu_1\ldots\mu_p}}_{\nu_1\ldots\nu_q}(x')= \Omega^{\Delta}(x){R^{\mu_1}}_{\lambda_1}(x)...{R^{\mu_p}}_{\lambda_p}(x){R_{\nu_1}}^{\rho_1}(x)...{R_{\nu_q}}^{\rho_q}(x) {{\hat{\phi}}^{\prime\lambda_1\ldots\lambda_p}}_{\rho_1\ldots\rho_q}(x)
\end{equation}
For a SCT, shifting the above tensor product leads to its original expression, since we are neglecting higher-order terms. For a dilation we get a product of Kronecker deltas. In the first case we obtain terms proportional to $x$, which cancel out when $x=0$. In the second case, the calculations are a carbon copy of the ones for a scalar field. 
The argument carries over for spinors. We can expand the exponential
\begin{equation}
\mathcal{D}\left[R(x)\right]=\exp\left(\frac{i}{2}\Theta_{\mu \nu}S^{\mu \nu}\right)=1+\text{\small terms directly proportional to $x$}.
\end{equation}
Since at the end we set $x=0$, we do not have to care about the additional terms. \\
We conclude that Definition \ref{def:definition2} implies Definition \ref{def:definition1}. 

\section{Derivation of the equivalence of the definitions: 1 implies 2}\label{sec:equivalence12}
To show that Definition \ref{def:definition1} implies Definition \ref{def:definition2}, one could start from the commutators between generators and field at $x=0$, with $\kappa_{\mu}=0$, and use the method of induced representation mentioned in Section \ref{sec:algebra} to obtain the full commutators \eqref{eq:commutators_fields}. At this point, one simply proceeds backwards through the steps outlined in the previous subsections until one arrives at the transformation \eqref{eq:def2}. Another possibility is the one proposed in \cite{Simmons-Duffin:2016gjk}. We revise it here with more detailed calculations. We start with the most general form of an infinitesimal conformal transformation, namely $\epsilon_{\mu}$ in Equation \eqref{eq:solution}. Using the commutators \eqref{eq:commutators_fields}, the corresponding infinitesimal variation of a field, $\delta\hat{\phi}^M(x)=-i\omega_aG^a(x)\hat{\phi}^M(x)$, is 
\begin{equation}
\begin{split}
    \delta\hat{\phi}^M&=-ia^{\mu}\left(-i\partial_{\mu}\right)\hat{\phi}^M-i\alpha\left(-i\Delta-ix^{\nu}\partial_{\nu}\right)\hat{\phi}^M\\&-i\frac{\omega_{\mu \nu}}{2}\left[i\left(x^{\mu}\partial^{\nu}-x^{\nu}\partial^{\mu}\right)+S^{\mu \nu}\right]\hat{\phi}^M
    \\&-ib^{\mu}\left(-2ix_{\mu}\Delta+ix^2\partial_{\mu}-2ix_{\mu}x^{\rho}\partial_{\rho}-2x^{\nu}S_{\mu \nu}\right)\hat{\phi}^M.
\end{split}
\end{equation}
We notice that 
\begin{equation}
\begin{gathered}
    \partial\cdot \epsilon=\partial_{\mu}\epsilon_{\nu}\eta^{\mu \nu}=\alpha d + 2(b\cdot x) d;\\
    \partial_{\mu}\epsilon_{\nu}-\partial_{\nu}\epsilon_{\mu}=2\omega_{\nu \mu} - 4b_{\nu}x_{\mu}-4b_{\mu}x_{\nu};\\
    \frac{1}{2}S_{\mu\nu}\partial^{[\mu}\epsilon^{\nu]}=\frac{1}{2}S_{\mu\nu}\frac{1}{2}\left(\partial^{\mu}\epsilon^{\nu}-\partial^{\nu}\epsilon^{\mu}\right)=\frac{1}{2}\left(-\omega^{\mu \nu}S_{\mu \nu}+4b^{\mu}x^{\nu}S_{\mu \nu}\right),
\end{gathered}
\end{equation}
where the last equality follows from the antisymmetry of $S_{\mu \nu}$ and $\omega_{\mu \nu}$.
Hence, we can write $\delta \phi^M$ as
\begin{equation}
    \delta\hat{\phi}^M(x)=-\left[\epsilon(x)\cdot \partial +\frac{\Delta}{d}\partial \cdot \epsilon(x)-\frac{i}{2}\partial^{[\mu}\epsilon^{\nu]}(x)S_{\mu \nu}\right]\hat{\phi}^M(x).
\end{equation}
In a quantum theory, a symmetry generated by $\hat{Q}_{\epsilon}$ is implemented by a 
unitary operator $\hat{U}=e^{-i\hat{Q}_{\epsilon}\epsilon}$ such that
\begin{equation}\label{eq:unitary_trafo}
    {\hat{\phi}}^{\prime M}(x')=\hat{U}\hat{\phi}^M(x)\hat{U}^{-1}.
\end{equation}
Using the Baker–Campbell–Hausdorff formula,
\begin{equation}
    e^{A}Be^{-A}=B+[A,B]+\frac{1}{2!}\left[\left[A,B\right],A\right]+...
\end{equation}
we can write equation \eqref{eq:unitary_trafo} as
\begin{equation}
    \hat{U}\hat{\phi}^M(x)\hat{U}^{-1}=\hat{\phi}^M(x)-i\epsilon\left[\hat{Q}_{\epsilon}, \hat{\phi}^M(x)\right]+O(\epsilon^2).
\end{equation}
The commutator is nothing but minus the variation of the field:
\begin{equation}
   i \epsilon  \left[\hat{Q}_{\epsilon}, \hat{\phi}^M(x)\right]=-\delta_{\epsilon}\hat{\phi}^M=\left[\epsilon(x)\cdot \partial +\frac{\Delta}{d}\partial \cdot \epsilon(x)-\frac{i}{2}\partial^{[\mu}\epsilon^{\nu]}(x)S_{\mu \nu}\right]\hat{\phi}^M(x).
\end{equation}
Hence
\begin{equation}
\begin{gathered}
    {\hat{\phi}}^{\prime M}(x')=\left(1-\epsilon\cdot \partial -\frac{\Delta}{d}\partial \cdot \epsilon+\frac{i}{2}\partial^{[\mu}\epsilon^{\nu]}S_{\mu \nu}\right)\hat{\phi}^M(x)=\\
   = \left(1-\frac{\Delta}{d}\partial \cdot \epsilon \right)\left(1+\frac{i}{2}\partial^{[\mu}\epsilon^{\nu]}(x)S_{\mu \nu}\right)\hat{\phi}^M(x)-\epsilon^{\mu}\partial_{\mu}\hat{\phi}^M(x)+O\left(\epsilon^2\right).
\end{gathered}
\end{equation}
For $\epsilon\to 0$ we recognize the first factor as the first term of the Taylor expansion of $\Omega^\Delta(x)$.  The second factor depends on which representation of the Lorentz group $\phi^M(x)$ is. For $\epsilon\to 0$ the term $\epsilon^{\mu}\partial_{\mu}\phi^M(x)$ simply vanishes. The final result is valid for an infinitesimal transformation. However, since a the conformal group is a Lie group, it is possible to achieve a finite transformation by composition of an infinite number of infinitesimal ones. Therefore, we can conclude that 
\begin{equation}
{\hat{\phi}}^{\prime M}(x')=\Omega^{\Delta}(x)\mathcal{D}{\left(R(x)\right)^M}_N\hat{\phi}^N(x),
\end{equation}
which is the content of Definition \ref{def:definition2}.

\section*{Acknowledgements}
The author thanks the Bethe Center for Theoretical Physics (BCTP) of the University of Bonn for its hospitality and Daniel Galviz for useful comments. This work was supported by the Bonn-Cologne Graduate School of Physics and Astronomy (BCGS).

\bibliographystyle{utphys}
\bibliography{references}

\providecommand{\href}[2]{#2}\begingroup\raggedright\begin{thebibliography}{10}

\bibitem{DiFrancesco:1997nk}
P.~Di~Francesco, P.~Mathieu, and D.~Senechal,
  \href{http://dx.doi.org/10.1007/978-1-4612-2256-9}{{\em {Conformal Field
  Theory}}}.
\newblock Graduate Texts in Contemporary Physics. Springer-Verlag, New York,
  1997.

\bibitem{Polchinski:1998rq}
J.~Polchinski, \href{http://dx.doi.org/10.1017/CBO9780511816079}{{\em {String
  theory. Vol. 1: An introduction to the bosonic string}}}.
\newblock Cambridge Monographs on Mathematical Physics. Cambridge University
  Press, 2007.

\bibitem{Blumenhagen:2009zz}
R.~Blumenhagen and E.~Plauschinn,
  \href{http://dx.doi.org/10.1007/978-3-642-00450-6}{{\em {Introduction to
  Conformal Field theory}: {With Applications to String Theory}}}, vol.~779 of
  {\em Springer Lecture Notes in Physics}.
\newblock Springer, 2009.

\bibitem{Ketov:1995yd}
S.~V. Ketov, {\em {Conformal field theory}}.
\newblock World Scientific, 1995.

\bibitem{Schellekens:1996tg}
A.~N. Schellekens, ``{Introduction to conformal field theory},'' {\em Fortsch.
  Phys.} {\bfseries 44} (1996) 605--705.

\bibitem{Schottenloher:2008zz}
M.~Schottenloher, \href{http://dx.doi.org/10.1007/978-3-540-68628-6}{{\em {A
  mathematical introduction to conformal field theory}}}, vol.~759 of {\em
  Springer Lecture Notes in Physics}.
\newblock Springer, 2008.

\bibitem{Cardy:2008jc}
J.~Cardy, ``{Conformal Field Theory and Statistical Mechanics},'' in {\em {Les
  Houches Summer School: Session 89: Exacts Methods in Low-Dimensional
  Statistical Physics and Quantum Computing}}.
\newblock 7, 2008.
\newblock \href{http://arxiv.org/abs/0807.3472}{{\ttfamily arXiv:0807.3472
  [cond-mat.stat-mech]}}.

\bibitem{Ginsparg:1988ui}
P.~H. Ginsparg, ``{Applied Conformal Field Theory},'' in {\em {Les Houches
  Summer School in Theoretical Physics: Fields, Strings, Critical Phenomena}}.
\newblock 1988.
\newblock \href{http://arxiv.org/abs/hep-th/9108028}{{\ttfamily
  arXiv:hep-th/9108028}}.

\bibitem{Dotsenko:1986ca}
V.~S. Dotsenko, ``{Lectures on Conformal Field Theory},'' {\em Adv. Studies in
  Pure Math.} {\bfseries 16} (1988) 123--170.

\bibitem{Qualls:2015qjb}
J.~D. Qualls, ``{Lectures on Conformal Field Theory},''
  \href{http://arxiv.org/abs/1511.04074}{{\ttfamily arXiv:1511.04074
  [hep-th]}}.

\bibitem{Rovai:2013gga}
A.~Rovai, ``{Introduction to conformal field theory},''
  \href{http://dx.doi.org/10.22323/1.195.0001}{{\em PoS} {\bfseries ModaveVIII}
  (2012) 001}.

\bibitem{Efthimiou:2000gz}
C.~J. Efthimiou and D.~A. Spector, ``{A Collection of exercises in
  two-dimensional physics. Part 1},''
  \href{http://arxiv.org/abs/hep-th/0003190}{{\ttfamily arXiv:hep-th/0003190}}.

\bibitem{Maldacena:1997re}
J.~M. Maldacena, ``{The Large N limit of superconformal field theories and
  supergravity},'' \href{http://dx.doi.org/10.1023/A:1026654312961}{{\em Adv.
  Theor. Math. Phys.} {\bfseries 2} (1998) 231--252},
  \href{http://arxiv.org/abs/hep-th/9711200}{{\ttfamily arXiv:hep-th/9711200}}.

\bibitem{Rattazzi:2008pe}
R.~Rattazzi, V.~S. Rychkov, E.~Tonni, and A.~Vichi, ``{Bounding scalar operator
  dimensions in 4D CFT},''
  \href{http://dx.doi.org/10.1088/1126-6708/2008/12/031}{{\em JHEP} {\bfseries
  12} (2008) 031}, \href{http://arxiv.org/abs/0807.0004}{{\ttfamily
  arXiv:0807.0004 [hep-th]}}.

\bibitem{Ammon:2015wua}
M.~Ammon and J.~Erdmenger, {\em {Gauge/Gravity Duality}: {Foundations and
  Applications}}.
\newblock Cambridge University Press, Cambridge, 2015.

\bibitem{Nastase:2015wjb}
H.~Nastase, {\em {Introduction to the ADS/CFT Correspondence}}.
\newblock Cambridge University Press, 2015.

\bibitem{Aharony:1999ti}
O.~Aharony, S.~S. Gubser, J.~M. Maldacena, H.~Ooguri, and Y.~Oz, ``{Large N
  field theories, string theory and gravity},''
  \href{http://dx.doi.org/10.1016/S0370-1573(99)00083-6}{{\em Phys. Rept.}
  {\bfseries 323} (2000) 183--386},
  \href{http://arxiv.org/abs/hep-th/9905111}{{\ttfamily arXiv:hep-th/9905111}}.

\bibitem{Zaffaroni:2000vh}
A.~Zaffaroni, ``{Introduction to the AdS-CFT correspondence},''
  \href{http://dx.doi.org/10.1088/0264-9381/17/17/306}{{\em Class. Quant.
  Grav.} {\bfseries 17} (2000) 3571--3597}.

\bibitem{Osborn:2013ConformalFT}
H.~Osborn, ``{Lectures on Conformal Field Theories in more than two
  dimensions},''
\newblock 2019.
\newblock
  \href{http://arxiv.org/abs/https://www.damtp.cam.ac.uk/user/ho/CFTNotes.pdf}{{\ttfamily
  https://www.damtp.cam.ac.uk/user/ho/CFTNotes.pdf}}.

\bibitem{Rychkov:2016iqz}
S.~Rychkov, \href{http://dx.doi.org/10.1007/978-3-319-43626-5}{{\em {EPFL
  Lectures on Conformal Field Theory in $D\geq3$ Dimensions}}}.
\newblock SpringerBriefs in Physics. Springer, 2016.
\newblock \href{http://arxiv.org/abs/1601.05000}{{\ttfamily arXiv:1601.05000
  [hep-th]}}.

\bibitem{Simmons-Duffin:2016gjk}
D.~Simmons-Duffin, \href{http://dx.doi.org/10.1142/9789813149441_0001}{``{The
  Conformal Bootstrap},''} in {\em {Theoretical Advanced Study Institute in
  Elementary Particle Physics}: {New Frontiers in Fields and Strings}}.
\newblock 2016.
\newblock \href{http://arxiv.org/abs/1602.07982}{{\ttfamily arXiv:1602.07982
  [hep-th]}}.

\bibitem{Mack:1969rr}
G.~Mack and A.~Salam, ``{Finite component field representations of the
  conformal group},''
  \href{http://dx.doi.org/10.1016/0003-4916(69)90278-4}{{\em Annals Phys.}
  {\bfseries 53} (1969) 174--202}.

\bibitem{Zee:2016fuk}
A.~Zee, {\em {Group Theory in a Nutshell for Physicists}}.
\newblock Princeton University Press, USA, 2016.

\bibitem{Poland:2018epd}
D.~Poland, S.~Rychkov, and A.~Vichi, ``{The Conformal Bootstrap: Theory,
  Numerical Techniques, and Applications},''
  \href{http://dx.doi.org/10.1103/RevModPhys.91.015002}{{\em Rev. Mod. Phys.}
  {\bfseries 91} (2019) 015002},
  \href{http://arxiv.org/abs/1805.04405}{{\ttfamily arXiv:1805.04405
  [hep-th]}}.

\bibitem{Weinberg:1995mt}
S.~Weinberg, {\em {The Quantum theory of fields. Vol. 1: Foundations}}.
\newblock Cambridge University Press, 6, 2005.

\end{thebibliography}\endgroup

\end{document}